\documentclass[journal,comsoc]{IEEEtran}
\ifCLASSINFOpdf
\else
\fi
\usepackage{amsmath}
\usepackage[cmintegrals]{newtxmath}
\usepackage[pdftex]{graphicx}
\usepackage{verbatim}
\usepackage{geometry}
\usepackage{makecell}
\usepackage{comment}
\usepackage{authblk}
\usepackage{amsmath,latexsym,url,amsfonts,setspace,xspace,color}

\usepackage{graphicx}

\newlength{\saveparindent}
\newlength{\saveparskip}
\newtheorem{thm}{Theorem} 
\newtheorem{lem}[thm]{Lemma}
\newtheorem{cor}[thm]{Corollary}
\newtheorem{propo}[thm]{Proposition}
\newtheorem{defn}[thm]{Definition}
\newtheorem{assm}[thm]{Assumption}
\newtheorem{clm}[thm]{Claim}
\newtheorem{rem}[thm]{Remark}

{\end{sl}\end{thm}}
{\end{sl}\end{lem}}
{\end{sl}\end{cor}}
{\end{sl}\end{propo}}
{\end{em}\end{defn}}
{\end{em}\end{assm}}
{\end{sl}\end{clm}}
{\end{em}\end{rem}}

\newcommand{\maybespace}{{\ifnum\springer=0{{\ }}\fi}}

\newcount\proofqeded
\newcount\proofended
\def\qed{ {\hspace{5pt}\rule[-1pt]{3pt}{9pt}}
\end{rm}\addtolength{\parskip}{-0pt}
\setlength{\parindent}{\saveparindent}
\global\advance\proofqeded by 1 }

 {\proofstart}%
 {\ifnum\proofqeded=\proofended\qed\fi \global\advance\proofended by 1
  \medskip}
\makeatletter
\def\proofstart{\@ifnextchar[{\@oprf}{\@nprf}}
\def\@oprf[#1]{\begin{rm}\protect\vspace{6pt}\noindent{\bf Proof of #1:\
}%
\addtolength{\parskip}{5pt}\setlength{\parindent}{0pt}}
\def\@nprf{\begin{rm}\protect\vspace{6pt}\noindent{\bf Proof:\ }%
\addtolength{\parskip}{5pt}\setlength{\parindent}{0pt}}

\newcounter{ctr}

\newcounter{ectr}

\newlength{\savejot}
\makeatletter

\newgeometry{vmargin={18mm}, hmargin={15mm,17mm}} 

\begin{document}
%
\title{A spatial-photonic Ising machine to solve the two-way number-partitioning problem}
%
%
%

\author[1*]{Vikram Ramesh}
\author[1]{Vighnesh Natarajan}
\author[1]{Anil Prabhakar}

\affil{\small{Department of Electrical Engineering, Indian Institute of Technology Madras, Chennai, India}}
\affil[*]{\small{Corresponding author: vikram.ramesh17@gmail.com}}

\maketitle

\begin{abstract}
We evaluate the performance of different algorithms in minimizing the Hamiltonian of a spatial-photonic Ising machine (SPIM). We then encode the number-partitioning problem on the SPIM  and adiabatically arrive at good solutions for the problem for over 16000 spins, with a time complexity that only scales linearly with problem size. Finally, we benchmark our machine performance against the classical solver, Gurobi, and also a D-Wave 5000+ quantum annealer. With just one spatial light modulator, and and adiabatic evolution scheme for the phase, our results surpass current state-of-the-art SPIMs. We reduce hardware costs, and can solve larger problems more efficiently.
\end{abstract}


%
\IEEEpeerreviewmaketitle

\section{Introduction}
NP-hard problems, by their very nature, require increasing (non-polynomial) computational resources as the number of variables increase.
Several applications such as vehicle routing, VLSI, or election voter distribution would benefit from efficient solution strategies to NP-hard problems such as the MAXCUT problem, or the number-partitioning problem (NPP) \cite{nphard}. 
Novel computing  architectures known as Ising machines promise to solve some of these problems by recasting them into Ising Hamiltonians, and then searching for a minimum energy ground state~\cite{Luc}.
The coherent Ising machine, implemented on optical parametric oscillators \cite{mcmahon, marandi, inagaki, inagaki2, bohm2} and optoelectronic or optical parametric oscillator networks, have used a recurrent feedback method to solve generic Ising models \cite{hart, bohm}. Degenerate cavity lasers have been employed in solving the phase retrieval problem \cite{nird1} and simulating the XY model Hamiltonian \cite{nird2}. Other optical Ising machines introduce nonlinearities through crystals \cite{kumar}.

Spatial-photonic Ising machines (SPIMs) are devices that embed a problem onto an optical wavefront, to then be solved in a Monte-Carlo like method. Spatial light modulators (SLMs), which have seen  widespread applications in laser beam shaping, optical trapping, vortex generation, etc. are employed in these machines. The programmability of SLMs allows for recurrent algorithms to iteratively correct the wavefront \cite{velle}, and their high pixel density allows them to process thousands of spins in parallel \cite{dpir,dpir2,dpir4,fang,kumar}. 

We have built an SPIM to minimize the Hamiltonian of a fully connected Ising model. In Sec. \ref{section:algos} we describe the experimental setup and then compare the performance of a genetic algorithm \cite{galletly} with the Metropolis-Hastings algorithm in minimizing the Hamiltonian. We then describe, in Sec. \ref{section:npp}, the theory and setup required to map a laser beam profile onto a Mattis model Hamiltonian \cite{matt} with programmable coupling constants. The Mattis model describes a spin glass system \cite{nishi}, and hence we demonstrate a noisy SPIM with a chosen algorithm. Using this, we solve the NPP for problem sizes upto over 16000 spins. In Sec. \ref{section:dwa} we benchmark the working of our system with other solvers of quadratic unconstrained binary objective (QUBO) problems. We show that adiabatically tuning the Hamiltonian provides solutions which are on par if not better than those achieved by the D-Wave system which can solve upto 121 spin systems. Additionally, the SPIM is capable of solving much larger problems than both the benchmark systems with linearly scaling time needed. Finally, we close with concluding remarks in Sec. \ref{section:conclusion}. Our solution strategy scales linearly at worst with the problem size on our current setup, and opens the door to parallel computing using photonic architectures.

\section{\raggedright Comparing convergence algorithms}

\begin{figure*}[!htbp]
\centering\includegraphics[width=0.7\linewidth]{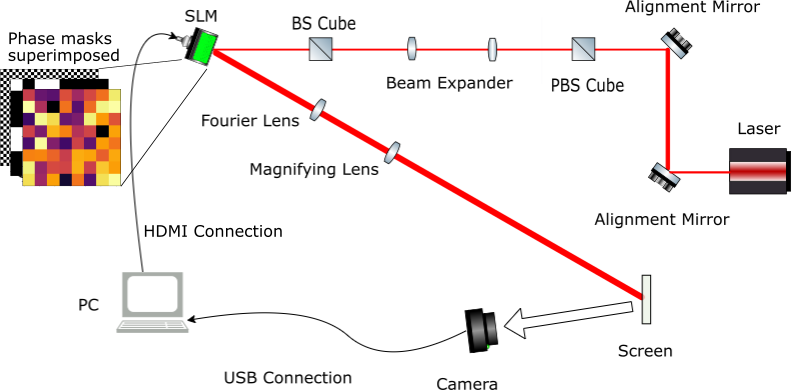}
\caption{Schematic of the experimental setup. The laser is linearly polarized by a PBS cube and using a beam expander, we fill out the active area of the SLM. The light reflected from the SLM undergoes a Fourier Transform after passing through a convex lens and subsequent magnification by a second convex lens increases the spatial resolution of the Fourier plane. At the top left of the figure the superimposed phase masks are indicated. These represent the binary checkerboard ($c_{j} = \pm 1$), phase mask of spins ($s_{j}=\pi/2$ or $3\pi/2$), and phase mask corresponding to the normalised numbers in the set ($\alpha_{j}=\cos^{-1}{\zeta_{j}}$). These phase masks are superimposed as $\theta_{j} = s_{j} + c_{j}\alpha_{j}$.}
\label{fig:setup}
\end{figure*}

\label{section:algos}
\subsection{Experimental Setup}
The experimental setup employed comprises a laser source, linear optical elements and an imaging system, as shown in a schematic given in Fig.~\ref{fig:setup}. A Gaussian laser beam from a 633 nm He-Ne laser is aligned with the help of 2 mirrors in kinematic mounts. The light is linearly polarized by a polarization beam splitter (PBS) aligned to the axis of the SLM. The beam profile is then expanded as it passes through 2 convex lenses. The intensity of the beam is reduced after passing through a beam splitter (BS). The resulting light is incident onto the screen of the SLM (Holoeye PLUTO-2-VIS-016). The wavefront at the SLM plane undergoes a Fourier Transform upon reflection by  passing through a convex lens of focal length 50 mm, and this Fourier object (captured at the back-focal plane of the lens) is then magnified using a second convex lens of the same focal length. Finally, the magnified object is captured on a screen and imaged by a CMOS camera (Basler acA2000-165um).

\subsection{Theory}
Let us take the electric field at the plane of the SLM to be $\Vec{E}(\Vec{r})$. Assume a polarized incident laser beam such that the incoming wavefront is
\begin{equation}
    \Vec{E}_{\text{in}}(\Vec{r}) = \sum_{j=0}^{N^{2}-1} \zeta_{j}\text{rect}_{j}(\Vec{r}) \hat{x},
\end{equation}
where $\zeta_{j}$ gives the complex amplitude of the electric field at the $j^{\text{th}}$ pixel of the SLM. We also assume that the active area of the SLM comprises $N \times N$ pixels of side length $L$, 
and that the laser spot is approximately a plane wave of constant amplitude $E_{0}$ over this active area and zero elsewhere. The $x$-component of the electric field at the SLM plane is written as
\begin{equation} \label{eq:eslm}
    E_{x}(\Vec{r}) = E_{0}\;\sum_{j=0}^{N^{2}-1} \phi_{j}\text{rect}_{j}(\Vec{r}),
\end{equation}
where $\text{rect}_{j}(\Vec{r})$ is the rectangular function  and $\phi_{j} = \exp{(\text{i} \theta_{j})}$. Here, $\theta_{j}$ is the phase delay imparted by the $j^{\text{th}}$ SLM pixel to the laser beam. The field given in (\ref{eq:eslm}) undergoes a Fourier Transform to become $\Tilde{E}_{x}(\Vec{k})$ at the camera plane \cite{goodman} and is given by
\begin{align}
    \Tilde{E}_{x}(\Vec{k})    &= E_{0} \int \;\sum_{j=0}^{N^{2}-1} \phi_{j}\text{rect}_{j}(\Vec{r}) \exp{(\text{i}\Vec{k}\cdot \Vec{r})} \text{d}^{2}r
\end{align}
The intensity $\Tilde{I}(\Vec{k})$ at the readout plane is hence given by:
\begin{multline}
    \left|\Tilde{E}_{x}(\Vec{k})\right|^{2}
    = L^{2}\text{sinc}^{2}\left(\frac{\Vec{k}\cdot \Vec{L}}{2}\right) \sum_{m,n=0}^{N^{2}-1} \zeta_{m}\zeta_{n}\phi_{m}\phi_{n}\\ \exp{\left[\text{i}\Vec{k}\cdot (\Vec{r_{m}}-\Vec{r_{n}})\right]}
\end{multline}
We now set a target intensity and define the cost function 
\begin{equation} \label{eq:cost}
    \text{Cost} = \sum_{x^{\prime},\;y^{\prime}} \left[I(x^{\prime},\;y^{\prime}) - I_{\text{Target}}(x^{\prime},\;y^{\prime})\right]^{2}
\end{equation}
where $x^\prime$ and $y^\prime$ are spatial coordinates in the camera plane. These represent the components of $\Vec{k} \propto x^{\prime} \hat{x} + y^{\prime} \hat{y}$. The constant of proportionality is given by $\frac{1}{\lambda f}$, where $\lambda$ is the wavelength of the laser and $f$ is the focal length of the Fourier lens. Since $I_{\text{Target}}(x^{\prime},\;y^{\prime})^{2} = \text{constant}$ and $\sum_{x^{\prime},\;y^{\prime}}I(x^{\prime},\;y^{\prime})^{2} \approx \text{constant}$ over iterations, we can take the Hamiltonian to be the cross product term such that

\begin{multline} \label{eq:algoham}
    H = -2L^{2}\int \text{sinc}^{2}\left(\frac{\Vec{k}\cdot \Vec{L}}{2}\right) \sum_{m,n=0}^{N^{2}-1} \zeta_{m}\zeta_{n}\phi_{m}\phi_{n}\\ \exp{[\text{i}\Vec{k}\cdot (\Vec{r_{m}}-\Vec{r_{n}})]} I_{\text{Target}}(k) \text{d}^{2}k
\end{multline}
If we consider a binary phase modulation by the SLM, i.e. $\phi_{m} = \pm 1$, then $H$ represents an all-to-all coupling in the Ising model, with coupling constants 
\begin{align} \label{eq:coupling}
    J_{mn} &= -2L^{2}   \mathcal{F}\left[\Tilde{I}_{\text{Target}}(k)\text{sinc}^{2}\left(\frac{\Vec{k}\cdot \Vec{L}}{2}\right)\right]  \zeta_{m}\zeta_{n} 
\end{align}
The coupling constant of two sites on the SLM depends on the chosen target intensity that we desire to settle to. To encode problems of our choice, we choose the target intensity to be a 2D delta function with peak at the central pixels of the camera, so that its effect after a Fourier transformation is constant, and we have a coupling term that depends on $\zeta_{m}\zeta_{n}$.

\subsection{Experimental Methods}

Using $256 \times 256$ pixels as the active area of the SLM, we group adjacent pixels as a spin. This aggregation is done to create sufficient contrast to be detected by the camera. We chose an active area of $256 \times 256$ pixels as it fills the laser spot on the SLM. Initializing a random spin distribution within the active area, we keep a constant binary checkerboard in the inactive area. Using this setup, we run iterative algorithms to move from the random spin distribution to the target distribution. Flipping $d=1$ spins within the active area at each iteration, we capture the resulting pattern with a CMOS camera.

The camera exposure time is set to give a maximum range for the intensity detection. The cost function of this pattern is then calculated with respect to the target image, as defined by (\ref{eq:cost}). At the $i^{\text{th}}$ iteration, $\text{Cost[i]}-\text{Cost[i-1]} = \Delta E$, and a Metropolis-Hastings (M-H) algorithm \cite{montecarlo} is run with an annealing schedule until the cost function converges to a minimum \cite{PATHRIA2011637, binder}.

\begin{figure}[htbp]
\centering\includegraphics[width=1.0\linewidth]{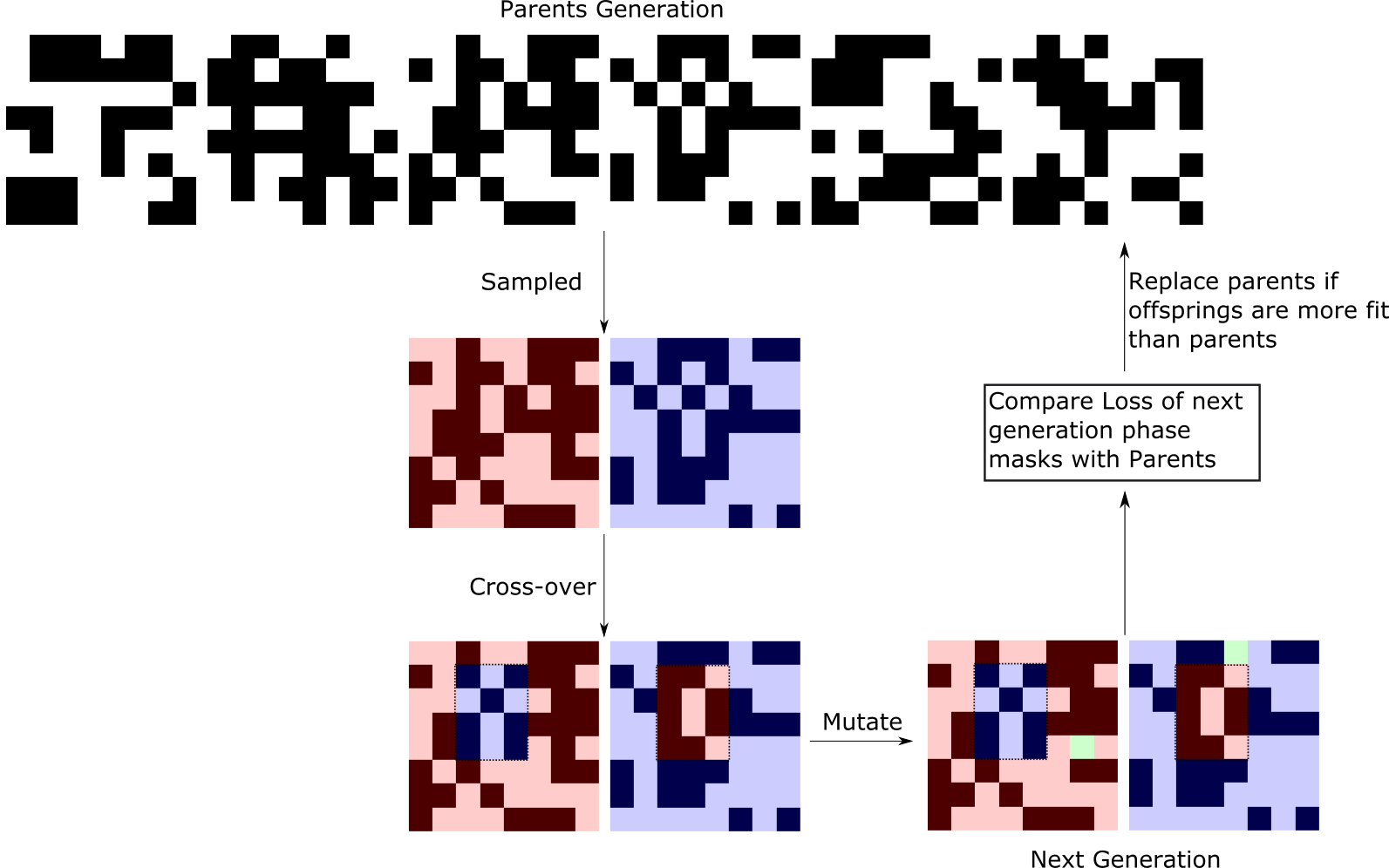}
\caption{An illustration of the iterative genetic algorithm for one generation of parents and offspring.}
\label{fig:GAFC}
\end{figure}

A traditional genetic algorithm (GA) was also used with this setup. Here the spin configurations are considered as bit strings representing the genes of an individual in a population \cite{galletly}. The cost function from (\ref{eq:cost}) was used here as a fitness function, and the population of spin configurations is evolved with a mutation probability of 0.05 per gene. The crossover from parents to offspring is 50-50 and randomly sampled from across the parents. A summary of this algorithm is given as a flowchart in Fig.~\ref{fig:GAFC}.

We quantify the limit or noise of the Ising machine as the average cost function for a series of images captured after uploading a checkerboard to the active area of the SLM, calculated with respect to a target image captured for the same checkerboard. The cost function used here is the same as that given in (\ref{eq:cost}).

\subsection{Results}

\begin{figure*}[htbp]
\centering\includegraphics[width=1.0\linewidth]{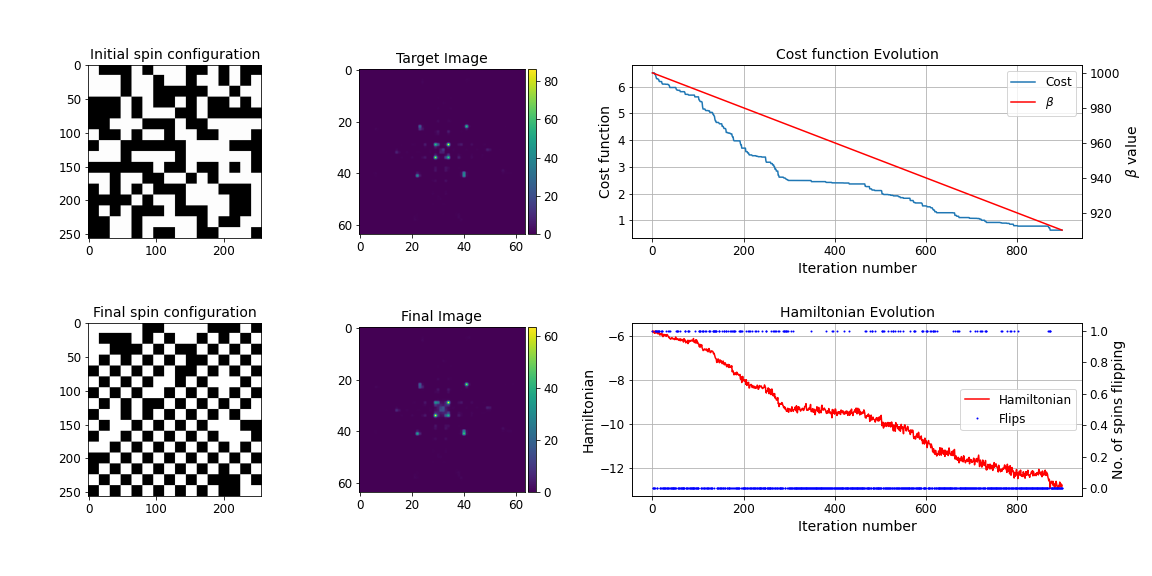}
\caption{Ising machine results for $16 \times 16$ spins. We flip $d = 1$ spins at each iteration. The checkerboard is almost entirely reproduced, as the cost function decreases to 0.04 of its initial value within around 800 iterations. The Hamiltonian function is correlated with the cost function defined in (\ref{eq:cost}) and similarly decreases. On the top right graph, the variable $\beta = \frac{1}{k_{B}T}$, where $k_{B} = 1.38 \times 10^{-23} \text{J}\text{K}^{-1}$ and $T$ is the temperature of the system.} 
\label{fig:MH}
\end{figure*}

\begin{figure*}[htbp]
\centering\includegraphics[width=1.0\linewidth]{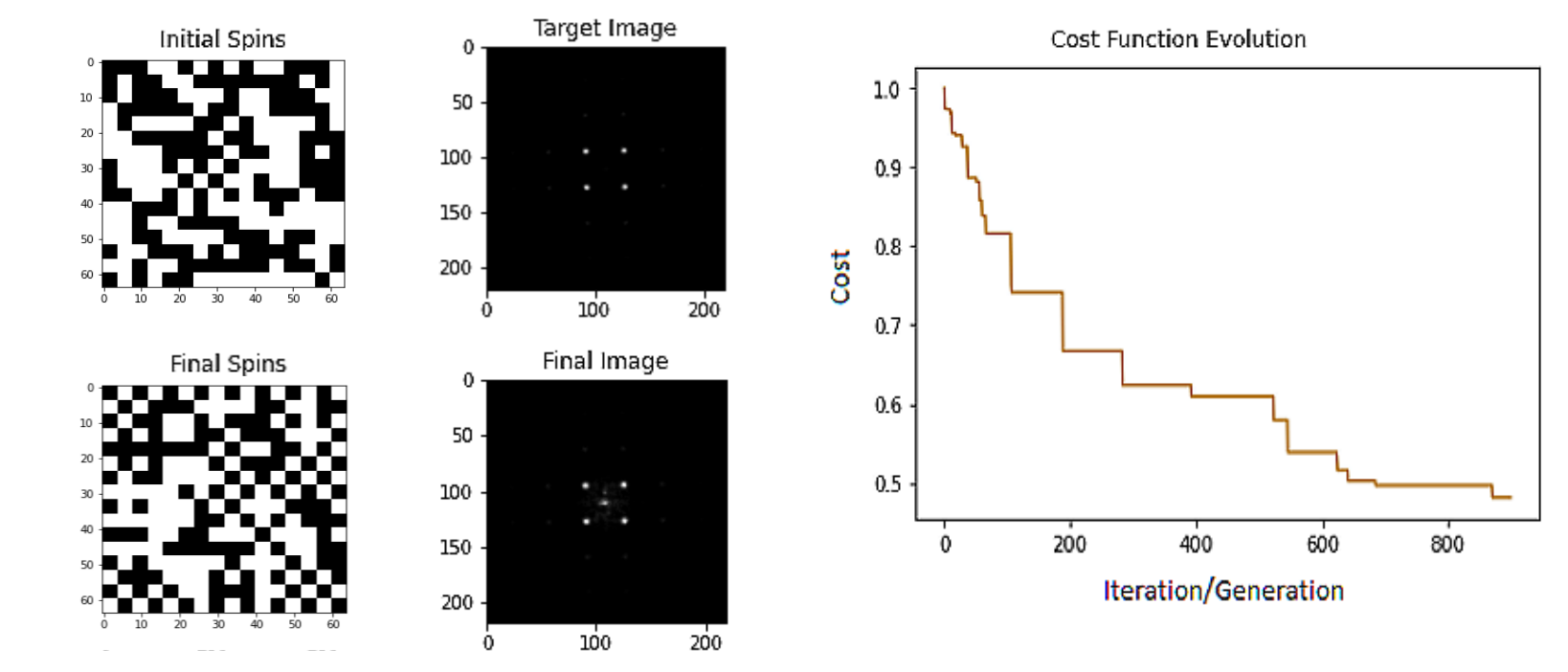}
\caption{Results when a genetic algorithm is run on the SPIM for a $16 \times 16$ spin lattice. The mutation rate is kept constant at 0.05 throughout the experiment. The cost (fitness) function only decreases to half of its initial value after 800 generations, and the checkerboard is only partially recovered.}
\label{fig:GA}
\end{figure*}

We see a steady decrease in the cost function for both algorithms as the experiment runs.  As shown in Fig.~\ref{fig:MH}, the cost function for the M-H algorithm decreases to within a $100^{\text{th}}$ of its initial value within 800 iterations for a $16 \times 16$ spin lattice. In the case of the GA, as shown in Fig.~\ref{fig:GA}, while we do see a decrease in the cost function, but it only reduces to half of its initial value after approximately 800 generations.
Thus, it takes twice as long to birth a new generation in the GA, when compared to an iteration of M-H. Add to this the insufficient decrease in the cost function, and we conclude that for a Mattis model, the M-H algorithm outperforms the GA, and is more robust to experimental noise generated by the optics. Hence, we choose the M-H algorithm to solve the NPP with our setup.

\section{\raggedright Number-partitioning problem} 
\label{section:npp}
\subsection{Theory}
\label{section:npptheory}
The Mattis spin glass\cite{matt} has a Hamiltonian of the form
\begin{equation} 
    H = \sum_{i,j}\zeta_{i}\zeta_{j}\sigma_{i}\sigma_{j},
    \label{eq:Mattis}
\end{equation}
where ${\sigma_{i}}$ is the binary spin at the $i^{\text{th}}$ lattice point and ${\zeta_{i}}$ is the amplitude contributing to the coupling between spins. This Hamiltonian can be exactly mapped to the objective function to be minimized to solve the NPP \cite{Luc}, subject to a proportionality constant. With the same experimental setup from Sec. \ref{section:algos}, we take the electric field at the plane of the SLM to be $\Vec{E}(\Vec{r})$, polarized along $\hat{x}$. Suppose the SLM active area comprises $N \times N$ pixels of side length $l$. Let $M \times M$ pixels be one spin, so we have  a lattice of $S \times S$ spins, with $N = MS$. As shown in Sec. \ref{section:algos}, and using the same approximations and notations, the x-component of the electric field at the camera plane is given by
\begin{align}
    \Tilde{E}_{x}(\Vec{k})    &= E_{0} \int \;\sum_{j=0}^{N^{2}-1} \phi_{j}\text{rect}_{j}(\Vec{r})\; \exp{(\text{i}\Vec{k}\cdot \Vec{r})} \text{d}^{2}r.
\end{align}
Putting $\Vec{k} = \Vec{0}$ gives the field at the origin of the readout plane as \cite{mendoza}
\begin{align}
    \Tilde{E}_{x}(\Vec{0}) & \propto -E_{0}l^{2} \sum_{j=0}^{N^{2}-1} \phi_{j}. 
\end{align}
where $\phi_{j} = \sigma_{j}e^{\text{i}(-1)^{j}\alpha_{j}}$. A similar idea was used in \cite{fang}. Here $\sigma_{j}$ gives the spin value and $\alpha_{j}=\cos^{-1}{\zeta_{j}}$, with $\zeta_{j}$ as the numbers normalized by dividing over the largest number in the set. These two terms are constant over an area of adjacent $M \times M$ pixels on the SLM, or within one spin. Hence, the electric field at the center of the camera plane is
\begin{equation}
     \Tilde{E}_{x}(\Vec{0}) \propto - E_{0}l^{2} \sum_{a=0}^{S^{2}-1} \sigma_{a} \left[e^{\text{i}\alpha_{a}}+  e^{-\text{i}\alpha_{a}}....M^{2}\text{ terms}\right]
\end{equation}
We choose $M$ to be even, which allows us to group pairs of exponentials and get,
\begin{align}
    \Tilde{E}_{x}(\Vec{0}) \propto - E_{0}l^{2}M^{2} \sum_{a=0}^{S^{2}-1} \sigma_{a}\cos{(\alpha_{a})}.
\end{align}
The intensity we detect at the center of the camera plane becomes
\begin{align}
    \Tilde{I}(\Vec{0}) & \propto   \sum_{m=0}^{S^{2}-1} \sum_{n=0}^{S^{2}-1} \sigma_{m} \Bar{\sigma_{n}} \cos{(\alpha_{m})}\cos{(\alpha_{n})}.
    \label{eq:I00}
\end{align}

\noindent Comparing (\ref{eq:Mattis}) and (\ref{eq:I00}), we observe that $\Tilde{I}(\vec{0})$ exactly maps onto $H$. 

\subsection{Experimental Methods}
With the experimental setup given in Fig.~\ref{fig:setup}, the laser intensity is recorded at the central $64 \times 64$ pixels on the image plane. The aim here is to minimize the total intensity captured at each iteration. A recurrent feedback loop is therefore setup between the camera and SLM through a Python programme on a desktop computer.

The light incident on the inactive area of the SLM also influences the readout. This was minimized by uploading a binary checkerboard of alternating up and down spins on the inactive area, minimized the  interference of the stray light at the center of the readout plane. The Hamiltonian given by (\ref{eq:Mattis}) was adiabatically changed from a problem instance of all equal numbers, to the desired problem instance, according to:
\begin{align}
    H(t) &\propto \sum_{m=0}^{S^{2}-1} \sum_{n=0}^{S^{2}-1} \sigma_{m} \sigma_{n} \cos{\left(\dfrac{t\theta_{m}}{T}\right)}\cos{\left(\dfrac{t\theta_{n}}{T}\right)}
\end{align}
The interpretation here is that the Hamiltonian initially represents a scenario where all the numbers to be partitioned are equal when $t = 0$, hence the checkerboard pattern for the spins is a ground state solution. These spins are rotated in phase on the SLM, which causes a change in the Hamiltonian to shift to represent the original Hamiltonian as we rotate the phase. When the value $t=T$, the Hamiltonian is the same as in (\ref{eq:I00}). Since the rotation here is cosine and not linear, the intermediate problem instances of the Hamiltonian are different from the target Hamiltonian. This phase imparted is to create the amplitudes that encode the number, and is superposed with the phase mask representing the spins as described in Sec. (\ref{section:npptheory}). The value of $t$ is not changed continuously, but in steps determined by the precision of the SLM. At each instance where the phases are changed, few iterations are given to let the SLM settle to the changed Hamiltonian, as can be observed in Fig.~\ref{fig:NPP}.

The SLM active area is chosen as $256 \times 256$ pixels or $512 \times 512$ pixels, depending on the size of the problem. Within this active area, we utilize the full analogue range of the SLM (8 bits) to simultaneously perform amplitude and phase modulation to get a Mattis model Hamiltonian at the readout plane. In our implementation, the range of problem instances depends on the size of the floating point variables sent to the SLM, which is 24 bits after the decimal place. For any number $\zeta = \cos\alpha$, we have $\text{d}\zeta = -\sin\alpha \; \text{d}\alpha$. Since the increment limit for the floating point is $\text{d}\alpha = 10^{-8}$, and $|\sin\alpha| \leq 1$, we get the upper limit as $\text{d}\zeta \leq 10^{-8}$. Hence, we can generate problem instances where each number has 8 significant digits at best.

Taking $N \times N$ adjacent pixels as a spin, a ground state is initialized, which for our initial Hamiltonian is any configuration that is 50\% phase $\pi/2$ and 50\% phase $3\pi/2$. The SLM is also divided into macropixels of size $2 \times 2$ pixels each. An additional phase of $(-1)^{j}\cos^{-1}\zeta_{m}$ is applied onto the $j^{\text{th}}$ macropixel of the SLM active area, where $m$ is the index of the spin.
Since the coupling constants are all positive in the problem instances we consider, adiabatically tuning them results in a dip in the cost function. The exposure time of the camera is chosen for each problem instance as the value required to just reach saturation of the intensity reading upon initializing a problem instance. This allows us to maximize the range of intensity values and hence the cost function change that the camera can detect, which leads to improved results.

At each stage in the adiabatic process \cite{dpir2}, a M-H algorithm is run by flipping $d$ spins at each iteration and taking a decision on whether to keep the flip based on the change in intensity, where the value of $d$ depends on the number of spins. Generally, we expect a lower value of $d$ to provide better convergence. We find an adiabatic solution by minimizing $H \propto I(\Vec{0})$ continuously as $\beta = (k_B T)^{-1}$ decreases, until a solution is reached.

This way to encode this problem onto a SPIM gives us an efficient means to get approximate solutions to the number partitioning problem

\subsection{Results}

\begin{figure*}[tbh]
\centering\includegraphics[width=0.9\linewidth]{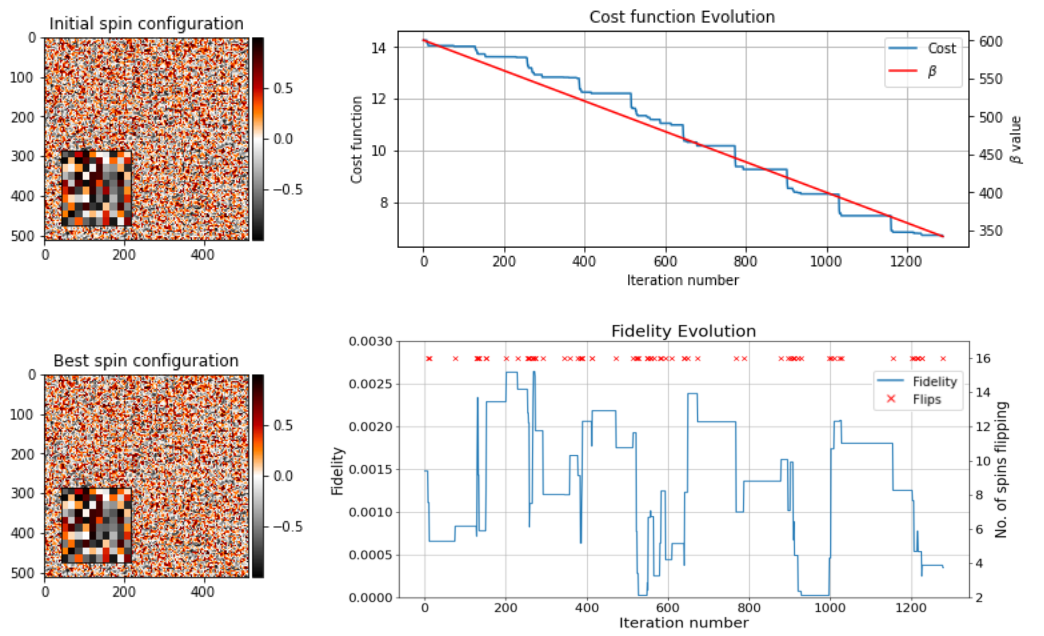}
\caption{Results for a problem instance of size 16384 spins. The periodic sudden dips in the cost function are a signature of the adiabatic tuning of the coupling constants. As shown in the plot on the bottom right of the figure, the fidelity escapes a local minima fairly easily, allowing us to sample a large energy landscape. Further, we can see that the number of accepted flips is lower than the number of rejected flips. Insets on the colourmaps shown in the left show an expanded view of a section of $10 \times 10$ spins.}
\label{fig:NPP}
\end{figure*}

\begin{figure}[tbh]
\centering\includegraphics[width=0.8\linewidth]{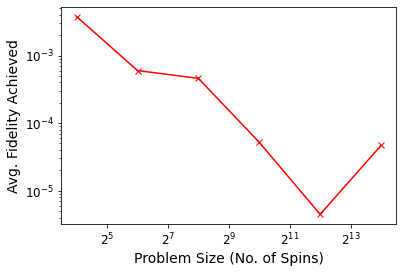}
\caption{Scaling of the solution quality with problem size. There is a mostly downward trend, which shows that solution quality increases on average with problem size. The experimental scheme is hence favourable in time complexity.}
\label{fig:fidelity}
\end{figure}

Adiabatically tuning the Mattis coupling coefficients improves the results, and the system consistently performs well for problem sizes ranging from 16 spins to 16384 spins. A sample plot for a problem instance of size 16384 spins is shown in Fig.~\ref{fig:NPP}. The plot at the top right of the figure shows the cost function decreasing throughout the experiment, with sharp dips whenever the coupling constants are changed. The quality of our solution is quantified by a fidelity,
\begin{equation}
    \eta = \left|\frac{\sum_{j} \zeta_{j}\sigma_{j}}{\sum_{j}\zeta_{j}}\right|
\end{equation}
\noindent
where $\zeta_{j}$ and $\sigma_{j}$ are the values of the $j^{\text{th}}$ number and spin. Squaring the fidelity gives:
\begin{equation}
    \eta^{2} = \left|\frac{\sum_{i,j} \zeta_{i}\sigma_{i}\zeta_{j}\sigma_{j}}{\sum_{i,j}\zeta_{i}\zeta_{j}}\right|
\end{equation}

The denominator in the above equation remains constant through all the iterations. Therefore, the square of the fidelity is proportional to the Mattis model Hamiltonian given in (\ref{eq:Mattis}) and by extension, the intensity at the center of the camera plane as well. The fidelity is thus positively correlated with the Hamiltonian throughout the energy landscape. In the plot shown on the bottom left of Fig.~\ref{fig:NPP}, we can see that the fidelity escapes a local minima quite easily, and hence we are able to sample a large solution space. This is partly due to the intrinsic intensity noise of the laser source, and partly due to the algorithm employed. The same plot also reveals that the number of accepted flips is a lot lower than the number of rejected flips. This is due to the Fourier transform, an all-to-all operation, which results in a spin flip causing only a tiny change to the detected intensity. A plot of the solution quality with problem size is shown in Fig.~\ref{fig:fidelity}, where the solution was averaged over several randomly generated problem instances. Unlike similar work done previously, we do not use the Hamming distance as a metric for the solution quantity \cite{dpir4}. The NPP typically has several approximate solutions which may be degenerate, and yet of starkly different spin configurations. The Hamming distance gives us information on how many spins need to be flipped so as to achieve the true ground state, but from an application point of view, the aim here is to benchmark the quality and utility of the achieved solution against other solvers.

\section{\raggedright Benchmarking the SPIM}
\label{section:dwa}

The performance of our SPIM in solving the NPP is compared with that of the D-Wave 5000+ qubit Advantage\_system1.1 Annealer (DWA) \cite{dwave} and the classical solver, Gurobi \cite{gurobi}, by running multiple problem instances on both devices. 
\begin{itemize}
    \item Table~\ref{tab:table1} compares the  best fidelity achieved over a single run for different problem instances of a fixed size. 
    \item Table~\ref{tab:table2} shows average fidelity achieved and its corresponding runtime, for the maximum tractable problem size.
\end{itemize}

\begin{table}[htb]
\centering
\caption{\bf Comparison of performance on different problem instances of size 64 spins. The best fidelity achieved for each solver during one run is shown below the solver name.\\}
\label{tab:table1}
\begin{tabular}{cccc}
\hline
\makecell{Problem. \\ No.} & \makecell{Gurobi} & \makecell{DWA} & SPIM \\
\hline
1 & 4.38E-05& 2.89E-05& 5.34E-04\\
2 & 3.58E-05& 1.78E-04& 1.28E-04\\
3 & 4.53E-05& 2.41E-03& 2.74E-04\\
4 & 1.07E-05& 1.27E-05& 1.47E-04\\
5 & 9.74E-05& 1.16E-04& 5.89E-04\\
\hline
\end{tabular}
\end{table}

\begin{table}[htb]
    \centering
    \caption{\bf Benchmarking the SPIM with the DWA and Gurobi. *The SPIM runtime indicated is the maximum required for all problem sizes. **The DWA runtime indicated includes the embedding time for 121 spins for different problem instances.\\}
    \resizebox{9cm}{!}{
    \begin{tabular}{cccc}
    \hline
     \makecell{Solver}    & \makecell{Max. Problem\\ Size}  & \makecell{Avg. Fidelity\\ for 64 spins} & Runtime\\
     \hline
      \makecell{SPIM} & 16384 & 6E-04 & \makecell{9 min.*}\\
      \makecell{DWA} & 121 & 5.49E-04 &  \makecell{$\sim$ 10 min.**}\\
      \makecell{Gurobi} & 1024 & 4.66E-05 & \makecell{$<$ 1 min. for \\64 spins\\ $\sim$ 10 min. for\\ 1024 spins}\\

      \hline
    \end{tabular}
    }
    \label{tab:table2}
\end{table}

We observe from Table \ref{tab:table1} that the SPIM achieves fidelity values of $\mathcal{O}(10^{-4})$ for different problem instances as opposed to the D-Wave device. Despite the fact that D-Wave has a 5000+ qubit system, the number of spins that can be embedded is capped at a $11 \times 11$ grid for the NPP. This is due to the large overhead in embedding a coupling between spins for a graph of density $100$\%. The runtime of the SPIM also scales favourably for larger problem instances when compared to other systems as we see in Table \ref{tab:table2}. For smaller problem sizes, up to 1024 spins, the performance of Gurobi is consistently better than both the D-Wave system and the SPIM and serves as a good benchmark to eventually achieve. However as the size of the problem gets larger, Gurobi is unable to fetch a solution, and the D-Wave annealer at present can't embed problem sizes of greater than a 11x11 grid = 121 spins due to limited connectivity. For larger problem sizes, the SPIM provides us a scalable method to solve the problem.

\section{Conclusion}
\label{section:conclusion}
Using minimal hardware, i.e. only one phase-modulation SLM with an additional magnifying lens, the fidelity values achieved show a marked improvement of 2 orders of magnitude over current state-of-the-art SPIMs. The SPIM performance was benchmarked against a D-Wave Annealer and Gurobi in solving the NPP. It was shown that the SPIM is favourable over DWA for its capacity to solve large problem sizes with similar performance for smaller problem sizes. The scope for improvement in hardware provides yet more promise for future applications. 

Presently, we are limited in speed by the response time of our SLM, which was measured to be about 150~ms before stabilizing (see Appendix).  The HDMI inputs to the SLM also limits the total iteration time to about 270~ms, since we must refresh the whole SLM for each spin update. Reductions in execution speed are possible with improvements to the hardware and firmware. We also expect to be able to run multiple instances of the problem, simultaneously, by partitioning the SLM and using different lenses to image each partition separately.

The accuracy of our mapping can be further improved with the addition of a passive diffractive-optical element to transform the Gaussian laser beam into a planar wavefront. Further, the algorithm is predicated on the adiabatic theorem of quantum mechanics, which deals with non-degenerate energy eigenstates, whereas in a typical instance of the NPP, several states may be degenerate. This could lead to potentially imperfect solutions. Another problem to tackle is the imperfect phase modulation by the SLM due to phase flicker, further enhanced by the oblique angle of incidence at the SLM plane. 

Currently, our hardware can only solve the Ising model based NPP. Looking forward, however, we hope to use Monte-Carlo methods similar to those mentioned in Sec. \ref{section:algos} to solve the XY and Potts models on our hardware \cite{xy,berloff}. Smart use of lenses would let us configure the connectivity of problems that can be embedded in the device, thus expanding the range of Hamiltonians that the SPIM can search. Future implementations of SPIMs may become more compact by replacing the Fourier lens with digital lenses programmed into the SLM \cite{Zhao:06}. Additionally, by employing fast adaptive optics technologies such as digital micromirror devices (DMDs) \cite{Goorden:14}, SPIMs may reach on/off speeds beyond the capability of SLMs. With the ongoing improvement of SLM specifications such as phase flicker and response time, the future of SPIMs holds exciting possibilities.


\textbf{Disclosures} The authors declare no conflicts of interest.

\textbf{Data availability} Data underlying the results presented in this paper are available under a Creative Common License~\cite{github}.





%

\appendix[SLM Response Time] \label{appendix:app}

The response time of the SLM was measured by uploading a binary grating and watching the changing diffraction pattern until it settles. It takes the array of liquid crystals approximately 150~ms to settle to the applied voltage, and this limits our iteration speed to less than 7~fps.

We begin by uploading a horizontal binary grating of $256 \times 256$ pixels onto the SLM screen and capturing the resulting diffraction pattern on a screen - which we then proceed to image with a camera. To characterize how fast the SLM uploads the full frame of phase values onto its screen, we re-initialize it with a random binary distribution of pixels - of phase values zero and $\pi$ - over the full frame of $1080 \times 1920$ pixels. The setup used is similar to that shown in Fig.~\ref{fig:setup}, but with a single phase mask. 

The camera region of interest (ROI) was set to be $1024 \times 1024$ pixels, in the free running mode. Upon executing the command to show a horizontal binary grating of $256 \times 256$ pixels onto the SLM, we subsequently capture a set of 30 images on the camera, and save the captured images into a video file. 

\begin{figure}[htb]
  \centering
  \includegraphics[width=0.47\linewidth]{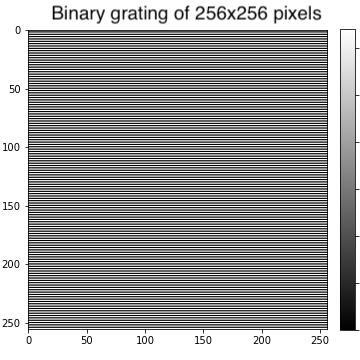} \hfill
    \includegraphics[width=0.5\linewidth]{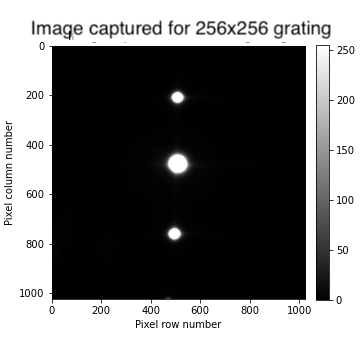}
  \caption{A binary grating of $256 \times 256$ pixels that is shown on the SLM screen (left). The image of the diffraction pattern generated by the grating in as captured by the camera (right).}
  \label{fig:grating256}
\end{figure}

We use two functions to aggregate the image data. The first function we use is the cost as given by (\ref{eq:cost}). Here the target intensity distribution is taken from the image shown on the right in Fig.~\ref{fig:grating256}. For the second function, we define 
\begin{equation}
    \label{eqn:energyresponse}
    \text{Energy} = \sum_{x,y} I(x,y),
\end{equation}
Both functions used are normalized to 1.

\begin{figure}[tbh]
    \centering
    \includegraphics[width=0.8\columnwidth]{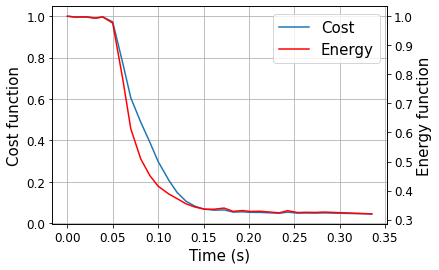}
    \caption{The response of the SLM is visualized by plotting the cost and energy functions with the time that the corresponding image was captured.}
    \label{fig:responseplot256}
\end{figure}

The image data is then post processed at each time step to produce two plots - energy and cost - as a function of the time at which the image was captured, as shown in Fig.~\ref{fig:responseplot256}.

The cost and energy functions in the plots shown in Fig.~\ref{fig:responseplot256} take time to start decreasing from their maximum values - and then settle within 0.2 seconds to a constant value after the grating has been fully uploaded. The origin of the time axis in the plots from Fig.~\ref{fig:responseplot256} is set to when the first image was captured i.e. the image acquisition command was executed after the command to show the grating onto the SLM screen. From these plots it is clear that a $256 \times 256$ binary grating takes at least 0.15 seconds to fully be shown onto the SLM screen. Hence, we are limited to working at less than 7~fps, where the spin patterns are sequentially uploaded to the SLM screen. Since the SLM size is much larger than the laser spot, running concurrent experiments on different sections of the SLM can speed up the process. 

\textbf{Acknowledgments} 
The authors thank Bagath Chandraprasad and Prof. Shanti Bhattacharya at the Indian Institute of Technology Madras for contributing their SLM, laser source and optical components. The authors would like to thank Prof. Sridhar Tayur and his team at Carnegie Mellon University for motivating the experiments and discussions on QUBOs. A. Prabhakar also thanks KLA for financial support.

\ifCLASSOPTIONcaptionsoff
  \newpage
\fi



%
\bibliographystyle{ieeetr}

\bibliography{bibliography}

%




\end{document}